\documentstyle[emulateapj,epsfig]{article}

\slugcomment{Ap. J. Vol. 583, in press}

\begin{document}

\title{Precursor Plerionic Activity and High Energy Gamma-Ray Emission
       in the Supranova Model of Gamma-Ray Bursts}

\author{Susumu Inoue$^{1,2,3}$, Dafne Guetta$^2$ and Franco Pacini$^2$}
\affil{$^1$Division of Theoretical Astrophysics,
       National Astronomical Observatory,
       2-21-1 Osawa, Mitaka, Tokyo, Japan 181-8588; inoue@th.nao.ac.jp\\
       $^2$Osservatorio Astrofisico di Arcetri,
       Largo E. Fermi, 5. 50125 Firenze, Italy;
       dafne@arcetri.astro.it, pacini@arcetri.astro.it\\
       $^3$Max-Planck-Institut f\"ur Astrophysik, Karl-Schwarzschild-Str. 1,
       85741 Garching, Germany; inouemu@mpa-garching.mpg.de}

\vspace{-4.0cm}
\begin{flushright}
{\small NAOJ-Th-Ap 2001, No.66}
\end{flushright}
\vspace{2.9cm}

\begin{abstract}
The supranova model of gamma-ray bursts (GRBs),
 in which the GRB event is preceded by a supernova (SN) explosion
 by a few months to years,
 has recently gained support from Fe line detections in the X-ray afterglows of some GRBs.
A crucial ingredient of this model yet to be studied
 is the fast-rotating pulsar
 that should be active during the time interval between the SN and the GRB,
 driving a powerful wind and a luminous plerionic nebula.
We discuss some observational consequences of this precursor plerion,
 which should provide important tests for the supranova model:
 1) the fragmentation of the outlying SN ejecta material
 by the plerion and its implications for Fe line emission;
 and 2) the effect of inverse Compton cooling and emission
 in the GRB external shock
 due to the plerion radiation field.
The plerion-induced inverse Compton emission
 can dominate in the GeV-TeV energy range during the afterglow,
 being detectable by GLAST from redshifts $z \lesssim 1.5$
 and by AGILE from redshifts $z \lesssim 0.5$, 
 and distinguishable from self-Compton emission by its spectrum and light curve.
The prospects for direct detection and identification
 of the precursor plerion emission
 are also briefly considered.
\end{abstract}


\section{Introduction}
\label{sec:intro}

In the currently standard interpretation of gamma-ray bursts (GRBs),
 the central engine gives rise to
 a highly relativistic outflow, the `fireball'.
Although the presence of relativistic outflows in GRBs
 has been amply demonstrated by multiwavelength observations of afterglows
 (Piran 1999, van Paradijs, Kouveliotou \& Wijers 2000, M\'esz\'aros 2001),
 the nature of the central engine itself is still a great mystery.  
One model of the central engine that has recently gained attention
 is the `supranova' model of Vietri \& Stella (1998, hereafter VS98),
 in which a massive star ends up in a supernova (SN),
 but the subsequent black hole formation and GRB event
 is delayed by some time $t_p$,
 typically expected to be a few months to years.
The SN is assumed to leave behind a rotationally-supported, `supramassive' neutron star
 (SMNS; hereafter also referred to simply as `pulsar')
 which then slowly shrinks by shedding angular momentum via a magnetospheric wind
 (and/or gravitational radiation; see \S\ref{sec:sum}).
After a spin-down time $t_p$ when roughly half its initial angular momentum has been lost,
 the configuration becomes unstable and collapses to a black hole,
 possibly with a surrounding disk, leading to the GRB proper.\footnote
{The numerical simulations of SMNS collapse to a black hole studied by
 Shibata, Baumgarte \& Shapiro (2000) do not support the formation of a debris disk
 that may power the GRB by accretion or external electromagnetic torques.
However, the GRB may also be triggered through the winding-up of strong magnetic fields
 due to differential rotation of the SMNS during the course of its collapse
 (e.g. Klu\'zniak \& Ruderman 1998).}
The model's major advantage
 lies in the potential realization of a very baryon-clean pre-GRB environment
 (mandatory for generating sufficiently relativistic fireballs):
 first the SN ejects the majority of the outlying mass of the progenitor star,
 and second the SMNS driven-wind
 effectively sweeps up remaining baryonic matter
 in the vicinity of the central object.
Support for this model comes from
 recent detections of strong Fe emission features in the X-ray afterglow spectra of some GRBs
 (Piro et al. 1999, Antonelli et al. 2001, Yoshida et al. 2001),
 particularly that of GRB991216 (Piro et al. 2000),
 as well as a transient Fe absorption feature
 in the prompt emission of GRB990705 (Amati et al. 2000),
 indicating surprisingly large amounts of Fe-rich material
 existing nearby, yet relatively removed from the GRB site.
Observational constraints on its location, quantity, density and velocity
 are compatible with the pre-ejected shell SN remnant (SNR) in the supranova model
 (Lazzati et al. 1999, Vietri et al. 2001,
 Lazzati et al. 2001),
 but may be difficult to accomodate in other models
 (see however Rees \& M\'esz\'aros 2000, M\'esz\'aros \& Rees 2001,
 B\"ottcher \& Fryer 2001, McLaughlin et al. 2002).

The precursor pulsar wind should be extremely powerful.
Before collapsing,
 the SMNS must inevitably expel a substantial fraction of its rotational energy $E_p$
 as a magnetically-driven wind,
 \begin{eqnarray}
 E_p = {G j M^2 \omega \over 2c}
     \simeq 2.4 \times 10^{53} {\rm erg} \ j_{0.6} M_{3}^2 \omega_4 ,
 \label{eqn:ep}
 \end{eqnarray}
 where $M = 3 M_3 {\rm M_\odot}$ and $\omega = 10^4 \omega_4 {\rm s^{-1}}$
 are the SMNS's mass and angular velocity,
 $j = 0.6 j_{0.6}$ is its angular momentum in units of $G M^2/c$,
 and the numbers are typical SMNS model values
 (Salgado et al. 1994, Cook, Shapiro \& Teukolsky 1994, VS98).
The spin-down time (i.e. the SN-GRB delay time) and corresponding wind luminosity
 can be estimated from the magnetic dipole formula (Pacini 1967, VS98)
 \begin{eqnarray}
 t_p = {3 G c^2 j M^2 \over B_*^2 R_*^6 \omega^3}
     \simeq 40 {\rm days} \ j_{0.6} M_3^2 R_{*,15}^{-6} \omega_4^{-3} B_{*,13}^{-2} ,
 \label{eqn:td}
 \end{eqnarray}
 and 
 \begin{eqnarray}
 L_p = E_p/t_p = {B_*^2 R_*^6 \omega^4 \over 6 c^3}
     \simeq 7.1 \times 10^{46} {\rm erg \ s^{-1}} R_{*,15}^6 \omega_4^4 B_{*,13}^2 ,
 \label{eqn:lp}
 \end{eqnarray}
 where $R_* = 15 R_{*,15} {\rm km}$ is a typical SMNS equatorial radius.
The surface magnetic field $B_* = 10^{13} B_{*,13} {\rm G}$ is unconstrained
 from model calculations and can be considered a free parameter.
Equivalently, we may take $t_p$ as the free parameter,
 and vary $L_p=E_p/t_p$ accordingly with $E_p$ fixed as in eq.\ref{eqn:ep};
 the fiducial value we choose below is $t_p \sim 120 {\rm days}$
 (implying $L_p \sim 2.3 \times 10^{46} {\rm erg \ s^{-1}}$)
 so as to be consistent with observations of GRB991216 (see \S\ref{sec:aftenv}).
During $t_p$, $L_p$ is expected to be relatively constant,
 and the wind should energize a plerionic nebula in the pre-GRB surroundings,
 a more compact yet much more luminous version of the Crab nebula.

The consequences of such a precursor plerion in the supranova scenario
 has not been considered previously,
 and this paper addresses some important dynamical and radiative effects it may induce,
 each providing important observational diagnostics for the supranova model.
We discuss
 the acceleration and fragmentation of the SN ejecta material
 by the plerion-SNR interaction
 and its implications for Fe line emission in \S\ref{sec:snr},
 and inverse Compton scattering of the ambient plerion radiation field
 in the GRB external shock
 and the resulting high-energy afterglow emission in \S\ref{sec:ec}.
A brief consideration of
 the direct detection and identification
 of the precursor plerion emission is given in \S\ref{sec:dir}.
We will assume a flat lambda cosmology
 with $\Omega_m=0.3$, $\Omega_{\Lambda}=0.7$ and $H_0=70 {\rm km s^{-1} Mpc^{-1}}$.

After this paper was submitted,
 we became aware of the work of K\"onigl and Granot (2002, hereafter KG02),
 who also investigated the properties of GRB afterglows
 occurring inside plerionic nebulae in the context of the supranova model.
They stressed the advantages of such a picture
 in realizing the relatively high magnetic fields and electron injection efficiencies
 required in the GRB blastwave to explain observed afterglows,
 and demonstrated this by constructing magnetohydrodynamical (MHD) models for the plerion
 under various assumptions.
Here we are interested in some characteristic observable effects
 caused by the precursor plerionic activity
 which are peculiar to the supranova model
 and through which the model
 may be tested by future observations.
The emphasis of this paper will be on the relevant radiative processes
 and their observational implications;
 we choose to keep the discussion of the plerion dynamics relatively simple
 and leave more detailed modeling of this aspect to future studies.
A study following our approach but with a more realistic treatment of the plerion
 has recently been carried out by Guetta \& Granot (2002).

\section{Dynamical Effects of the Plerion on the Supernova Remnant}
\label{sec:snr}

An exemplary case of the dynamical interaction between a plerion and a SNR
 can be seen in the well-studied Crab nebula,
 which is known to be accelerating and fragmenting the surrounding SNR,
 resulting in the prominent optical filaments instead of a clear shell
 (e.g. Davidson \& Fesen 1985, Hester et al. 1996).
In this respect, the powerful supranova plerion can
 be even more effective than the Crab.
We model the plerion in a simple way
 following Pacini \& Salvati (1973, hereafter PS73;
 see also Bandiera, Pacini \& Salvati 1984, Chevalier \& Reynolds 1984, Chevalier 2000),
 considering a homogeneous, spherical bubble
 into which energy is injected at a constant rate $L_p = 10^{46}L_{p,46} {\rm erg \ s^{-1}}$,
 a fraction $\xi_B=0.5\xi_{B,0.5}$ going into magnetic field
 and the rest $\xi_e=1-\xi_B=0.5\xi_{e,0.5}$ into relativistic electrons.
The electrons here are mostly radiative (see \S\ref{sec:ec}),
 and we neglect their pressure for simplicity.
As the plerionic bubble initially plows through
 the expanding core and envelope of the progenitor star,
 it should accelerate the swept-up ejecta material
 (Chevalier 1977, Chevalier \& Fransson 1992, KG02).
If a fair fraction of the total plerion energy $E_p \simeq 10^{53} {\rm erg}$
 can be conveyed to SN ejecta of several ${\rm M_\odot}$,
 their final attained velocities could reach $v_s \sim 0.05-0.1c$,
 consistent with those inferred from
 the observed width of the Fe line in GRB991216 (Piro et al. 2000).

As in the Crab,
 the SN ejecta should not entirely remain as a spherical shell during this acceleration phase
 due to Rayleigh-Taylor (RT) instabilities operating at the plerion-SNR interface.
The growth timescale of the RT instability on a spatial scale $R$
 is $t_{RT} \sim (R/\ddot R)^{1/2}$,
 and since $\ddot R \sim 4\pi R^2 p_B/M_s$,
 $p_B = B_p^2/8\pi$ being the plerion magnetic pressure,
 its ratio to the expansion timescale $t_{exp} \sim R/v_s$ is
 \begin{eqnarray}
 {t_{RT} \over t_{exp}} &\sim& \left({2 M_s v_s^2 \over B_p^2 R^3}\right)^{1/2}
         \sim \left({4 E_s \over 3 \xi_B E_p}\right)^{1/2} \nonumber\\
         &\simeq& 0.39 E_{s,51}^{1/2} (\xi_{B,0.5} L_{p,46})^{-1/2} t_{10}^{-1/2}
 \label{eqn:trt}
 \end{eqnarray}
 (Bandiera, Pacini \& Salvati 1983),
 where $M_s$, $v_s$ and  $E_s = M_s v_s^2/2 = 10^{51} E_{s,51} {\rm erg}$
 are the SNR ejecta mass, velocity and kinetic energy,
 and $t_{10}$ is time after the SN in units of 10 days.
In the case of the Crab, $t_{RT} / t_{exp} \gtrsim 1$
 and the RT instability is currently setting in,
 consistent with other lines of evidence
 (Hester et al. 1996, Sankrit \& Hester 1997).
However, it
 can potentially develop much more rapidly for the supranova plerion.

The final outcome of the RT instability is not easy to predict,
 as it will depend on the instability's non-linear behavior
 as well as other processes such as radiative cooling and thermal conduction.
We refer to the work of Jun (1998) as a guideline,
 who carried out 2-dimensional numerical simulations
 of the plerion-SNR interaction and associated RT instabilities,
 choosing Crab-like parameters for $L_p$ and $t$
 and neglecting cooling.
His results demonstrate that
 the RT instability leads to dramatic effects at late times,
 strongly disrupting the SNR shell and transforming it into pronounced RT ``fingers''
 which carry the majority of the SN ejecta mass and kinetic energy.
To be identified with the Crab's optical filaments,
 these fingers protrude inward from the outer ejecta shell and end in dense ``heads''
 with slower expansion velocities than the plerion,
 which can eventually become decoupled from the shell.
In Jun's simulations,
 the density of the heads was on average $\sim 10$ times that of the rest of the SN ejecta,
 and the radial extent of the fingers was $\sim 0.2$ times that of the shell radius.
However, as mentioned in his discussion,
 the inclusion of radiative cooling should increase the density contrast of the heads,
 and both this and the radial extent of the fingers should grow with time.
These quantities could then be much larger for the supranova plerion, 
 since the relative timescale of the instability here might be much shorter (eq.\ref{eqn:trt}).
Although a firm conclusion warrants high resolution simulations of the relevant conditions,
 it is not implausible
 that the plerion can
 effectively shred a substantinal portion of the SN ejecta into condensed fragments
 and engulf them during its lifetime $t_p$.
Note that Arons (2002), in a somewhat different context,
 has recently discussed the issue
 of SNR fragmentation by a powerful pulsar wind in greater depth.

If this is indeed the case
 (see \S 5 for comments when this is not),
 there are
 important implications for the Fe line emission in afterglows.
The simplest explanation for the lines is Fe K$\alpha$ multiple recombination radiation
 from Fe-rich matter photoionized by the X-ray afterglow continuum on timescales of days
 (Lazzati et al. 1999, Weth et al. 2000, Paerels et al. 2000).
Consistency with observations require
 that the material to be illuminated is very dense
 and possesses a large covering factor,
 and yet that the GRB line of sight
 is devoid of any such matter
 so that the GRB blastwave will not be decelerated too quickly (Lazzati et al. 1999).
The filamentary fragmentation of the SNR
 may naturally account for such a geometry.

The condensation of the SNR matter may also enhance the line emissivity.
The emission rate of line photons through recombination
 when a sufficient photoionizing flux irradiates
 material with electron density $n_e = 10^9 n_{e,9} {\rm cm^{-3}}$
 and temperature $T = 10^7 T_7 K$ 
 containing an Fe mass $M_{\rm Fe} = 0.1 M_{\rm Fe, 0.1} {\rm M_\odot}$
 is
 \begin{eqnarray}
 \dot N_{line} = N_{\rm Fe} / t_{rec}
               \simeq 1.5 \times 10^{52} {\rm s^{-1}} M_{\rm Fe, 0.1} n_{e,9} T_7^{-3/4}
 \label{eqn:nl}
 \end{eqnarray}
 where $N_{\rm Fe}$ is the total number of Fe atoms
 and $t_{rec} \simeq 127 {\rm s} \ T_7^{3/4} n_{e,9}^{-1}$ is the recombination time
 for H-like Fe (Paerels et al. 2000, Vietri et al. 2001).
By the time of the GRB at $t_p = 120 t_{p,120} {\rm days}$ (see \S\ref{sec:ec}),
 a SNR of $M_s = 10 M_{s,10} {\rm M_\odot}$ expanding at $v_s = 0.1 v_{s,0.1} c$
 will have reached a radius $R_s \simeq 3.1 \times 10^{16} v_{s,0.1} t_{p,120} {\rm cm}$,
 and if distributed in a spherical shell of width $\Delta R_s \sim 0.1 R_s$,
 its electron density would be
 \begin{eqnarray}
 n_{e,s} \simeq 5.3 \times 10^8 M_{s,10} (v_{s,0.1} t_{p,120})^{-3} {\rm cm^{-3}} .
 \label{eqn:nes}
 \end{eqnarray}
Typically observed line fluxes can be obtained for $M_{\rm Fe, 0.1} \gtrsim 4$ (Eq.\ref{eqn:nl}),
 but only if the entire spherical shell is irradiated and emits line photons efficiently.
The GRB blast wave must then be isotropic and no beaming is allowed
 (Piro et al. 2000; see however KG02),
 which is incongruous with recent evidence to the contrary
 (e.g. Frail et al. 2001, Panaitescu \& Kumar 2002).
However, if the SNR ejecta can be condensed into filaments of much higher density,
 $t_{rec}$ is shortened,
 and irradiating a smaller fraction of material may suffice to produce the observed lines.
For comparison, the density of the the Crab's optical filaments
 averaged over the volume of the whole nebula is $n \sim 10 {\rm cm^{-3}}$,
 whereas the actual density of the individual filaments is $n \sim 10^{3} {\rm cm^{-3}}$
 (Davidson \& Fesen 1985).
Although detailed numerical modeling is necessary for quantitative predictions,
 we may speculate that
 the density contrast could be even higher for the supranova plerion
 due to the instability being more developed (see above),
 and beaming fractions of order $\sim 0.01$ may be entirely compatible
 with the observed Fe lines.
Note that
 in explaining the transient Fe absorption feature in GRB990705,
 Lazzati et al. (2001) have inferred the presence of
 clumpy SNR ejecta material with density contrasts of order 100-1000
 surrounding the burst site (see also B\"ottcher, Dermer \& Fryer 2002).
The observed line energy in GRB991216 can also be consistent with beaming
 if identified with blueshifted emission from He-like Fe
 (for which the rest frame energy is 6.7 keV,
 and $t_{rec}$ is similar to H-like Fe at somewhat lower $T$)
 in material approaching toward the observer with $v_s \sim 0.1c$
 (Ballantyne \& Ramirez-Ruiz 2001).

We mention that even when circumstances conducive to Fe line emission cannot be realized,
 the action of the plerion for $t_p \gtrsim 10^3 {\rm s}$
 may expel much of the baryons surrounding the GRB site out to radii $\gtrsim 10^{13} {\rm cm}$
 and clear the path for the ensuing fireball,
 which is a great advantage of the supranova model.

\section{External Compton Cooling and Emission due to the Plerion Radiation Field}
\label{sec:ec}

Another important implication of the precursor plerion in the supranova model
 is that the GRB should be exploding into a `radiation-rich' environment,
 i.e. into the luminous radiation field of the plerion,
 which can constitute a large fraction of the pulsar wind luminosity.
Such `external' photons
 impinging into the relativistically expanding GRB shock from outside
 would be seen highly Doppler-boosted in the shock comoving frame,
 and act as efficient seeds for inverse Compton scattering
 (referred to as external Comptonization, or EC),
 with important observational consequences for the afterglow emission.
\footnote{
EC processes have been extensively discussed in relation to blazars; see e.g. Sikora 1997.
For previous work on EC processes in GRB shocks, see e.g. B\"ottcher \& Dermer 1999.
}

Since our focus here is on the possible manisfestations
 of the EC emission process in the supranova model,
 we do not attempt a detailed physical modeling of the the plerion dynamics,
 employing instead a simplified description for illustrative purposes.
As discussed by KG02, such modeling would entail various uncertainties
 and requires a dedicated study by itself, which is out of our current scope.
Once a suitable dynamical model of the plerion can be constructed,
 the formulation of this paper may be used
 to make more reliable predictions of the afterglow emission.
Alternatively, if an afterglow EC emission component is actually observed in the future,
 we may interpret the data with our formulation
 so that the unknown properties of the supranova plerion can be probed.

\subsection{Plerion Emission}
\label{sec:ple}

We continue the discussion of the plerion
 as a roughly homogeneous, spherical bubble with constant energy injection.
Following the arguments of the preceding section,
 it is postulated that the plerion can first penetrate effectively
 through the expanding SN ejecta, shredding and entraining the ejecta material
 within a relatively short time.
Thereafter, the plerion can be considered to expand
 into the ambient medium, here taken to be uniform for brevity.
The consequences of the dense, entrained baryonic clumps 
 for the global dynamics of the plerion is unclear,
 but lacking a realistic model, we do not delve into this complication
 and make simplified assumptions regarding its evolution.
If the SNR filaments are expanding at velocities of order $v_s \sim 0.1c$
 (as implied by the observed Fe line width in GRB991216),
 the expansion velocity of the plerionic bubble $v_p$
 may be several times larger than this,
 judging from the observed expansion velocities in the Crab
 as well as the numerical simulations of Jun (1998).
As long as the ejecta material is not dynamically well-coupled to the plerion as a whole,
 we may also estimate $v_p$ from the relativistic blastwave solution of Blandford \& McKee (1976)
 in the case of steady, adiabatic energy injection into a uniform medium,
 which gives
 $\Gamma_p \sim ({k L_p / \rho_{\rm I} c^5})^{1/4} t_p^{-1/2}
          \simeq 2.9 L_{p,46}^{1/4} n_{\rm I,0}^{-1/4} t_{p,120}^{-1/2}$,
 where $\Gamma_p=(1-v_p^2/c^2)^{-1/2}$ is the expansion bulk Lorentz factor,
 $n_{\rm I} = \rho_{\rm I}/m_p c^2 = n_{\rm I,0} {\rm cm^{-3}}$ is the ambient density
 and $k$ is a numerical factor of order unity.
\footnote{
This formula strictly applies only in the ultrarelativistic limit ($\Gamma_p \gg 1$, $v_p \sim c$);
 in the nonrelativistic limit ($\Gamma_p \sim 1$, $v_p \ll c$),
 the deceleration goes as $v_p \propto t_p^{-2/5}$ and $R_p \propto t_p^{3/5}$.
}
A mildly relativistic expansion is then inferred
 for ages $t_p \sim 10^2-10^3 {\rm days}$
 and the values of $L_p$ and $n_{\rm I}$ considered here.
We assume, at least for the conditions of our interest,
 that the plerion has been expanding at a nearly constant velocity of $v_p \sim c$,
 so that its radius $R_p \sim c t_p \simeq 3.1 \times 10^{17} t_{p,120} {\rm cm}$.

The luminosity and spectrum emitted by the plerion
 may be evaluated following PS73.
The magnetic field inside the plerion (taken to be isotropically tangled) is
 $B_p = \left(3 \xi_B E_p / R_p^3\right)^{1/2}
      \simeq 3.5 {\rm G} \xi_{B,0.5}^{1/2} t_{p,120}^{-3/2}$,
 with the value of $E_p$ fixed as in eq.\ref{eqn:ep}.
Relativistic electrons/positrons (hereafter simply 'electrons')
 are injected into the plerion with a power of $\xi_e L_p$,
 having an isotropic, power-law energy distribution $dn/d\gamma^p \propto (\gamma^p)^{-s}$
 in the Lorentz factor range $\gamma^p_m \le \gamma^p \le \gamma^p_M$.
We fiducially take $\xi_B=\xi_e=0.5$, $s=2$ and $\gamma^p_m=1$,
 parameters which lead to consistent fits
 when similar models are applied
 to the multiwavelength spectra of the Crab and other known plerionic nebulae
 (PS73, Bandiera et al. 1984, Amato et al. 2000, Chevalier 2000).
The maximum Lorentz factor $\gamma^p_M$ is set
 by equating the synchrotron cooling time
 $t^p_{sy} \sim 6 \pi m_e c / \sigma_T B_p^2 \gamma_p$
 with the acceleration time, here assumed to be
 $t^p_{acc} \sim 2\pi \gamma_p m_e c / e B_p$
 (consistent with Fermi acceleration at relativistic shocks),
 so that $\gamma^p_M \sim (3e / \sigma_T B_p)^{1/2}$
 and the corresponding maximum synchrotron emission frequency
 $\nu^p_M \sim 3 e^2 / \pi m_e c \sigma_T \simeq 1.2 \times 10^{22} {\rm Hz}$.
\footnote{
With our fiducial choice of $\xi_B=\xi_e=0.5$,
 synchrotron-self-Compton cooling is at most comparable
 to synchrotron cooling, so this is neglected for simplicity.
}
The distribution should develop a break at the Lorentz factor $\gamma^p_c$
 where $t^p_{sy}$ equals the adiabatic expansion time
 $t^p_{ad} \sim R_p/v_p \sim t_p$, leading to $\gamma^p_c \sim 6\pi m_e c / \sigma_T B_p^2 t_p$.
Our fiducial parameters give $\gamma^p_c \sim 6$,
 implying that injected electrons of all but the lowest energies
 are radiative and cool within $t^p_{ad}$;
 the cooled distribution has index $s+1$
 and emission spectrum of energy index $\alpha_p \sim s/2=1$.
The specific flux at the surface of the plerion is
 \begin{eqnarray}
 f^p_{\nu^p} \sim {\xi_e L_p (\nu^p)^{-1} \over 8\pi R_p^2 \ln(\gamma^p_M/\gamma^p_m)} .
 \label{eqn:fnup}
 \end{eqnarray}
Synchrotron self-absorption will
 cause an effective minimum cutoff in the spectrum at a frequency
 \begin{eqnarray}
 \nu^p_a
  &=& {e \over 2\pi m_e c} \left({\pi \sqrt{3\pi} \over 4} 3^{s+1 \over 2} f(s+1) {e K_c R_p}\right)^{2 \over s+5} B_p^{s+3 \over s+5} \nonumber \\
  &\simeq& 2.4 \times 10^{11} {\rm Hz} \xi_{e,0.5}^{2/7} \xi_{B,0.5}^{1/14} t_{p,120}^{-15/14} ,
 \label{eqn:nupa}
 \end{eqnarray}
 where $f(s+1)$ is a factor of order unity weakly depending on $s$ (e.g. Rybicki \& Lightman 1979),
 and
 \begin{eqnarray}
 K_c \equiv {dn_c \over d\gamma^p} / (\gamma^p)^{-s-1} \sim {6 \xi_e \over \sigma_T c \xi_B t_p \ln(\gamma^p_M/\gamma^p_m)}
 \label{eqn:kc}
 \end{eqnarray}
 is the normalization factor of the cooled electron distribution (PS73).
The expressions in the last lines of eqs.\ref{eqn:nupa} and \ref{eqn:kc}
 are for our fiducial choice of $s=2$. 
As the flux falls steeply below $\nu^p_a$,
 the overall plerion spectrum
 is that of eq.\ref{eqn:fnup} between $\nu^p_a$ and $\nu^p_M$,
 with constant luminosity per logarithmic frequency interval,
 and the total emitted luminosity is of order $\xi_e L_p$.
\footnote{
Any flux above $\nu^p_M$ which may arise from self-Compton emission is irrelevant
 for electrons in the GRB shock due to Klein-Nishina effects; see below.
}
Note that
 some quantities regarding the plerion contain subscripts or superscripts $p$
 in order to avoid confusion with those for the GRB blastwave.

We mention that the plerion should also drive a forward shock into the ambient medium,
 whereby the associated synchrotron radiation
 may give rise to an additional emission component.
Since the magnetic field and the electron injection efficiency in the shocked external medium
 are expected to be lower than that inside the plerion,
 such emission should be of subdominant luminosity
 appearing at frequencies below $\nu^p_a$.
EC scattering of this component by GRB blastwave electrons
 is unimportant for the observed afterglow,
 as it is likely to be masked by the synchrotron and SSC radiation
 occurring in the pertinent frequency range (see below);
 however the emission at radio frequencies may play a role
 in the direct detectability of the plerion before the GRB (\S 4).
A proper account of this plerion forward shock emission
 mandates a more detailed description of the plerion expansion,
 so in this study we choose to neglect this component
 and restrict ourselves to emission from the plerion interior.

\subsection{GRB Afterglow Emission}
\label{sec:aftemi}

At time $t_p$ after the SN, the pulsar collapses and the GRB goes off,
 sending a fireball and relativistic blastwave into the plerion.
We may distinguish two different situations
 for how the deceleration of the GRB blastwave, and hence the afterglow, initiates.
As in conventional discussions of afterglows,
 the deceleration may begin when it has swept up enough outlying baryonic material,
 which may occur if the (probably collimated) fireball in our line of sight happens to strike
 the denser parts of the entrained SN ejecta fragments inside the plerion.
Alternatively, the deceleration may result from
 the blastwave accumulating enough inertia
 in relativistic particles and magnetic field of the pulsar wind material, as discussed by KG02.
To describe the afterglow in this case,
 KG02 showed that standard formulae (e.g. Sari, Piran \& Narayan 1998) can still be used
 if one replaces parameters such as the ambient medium density
 and electron and magnetic energy density fractions
 by equivalent quantities related to the plerion properties (see below).
In considering either case, we will assume for simplicity
 that the decelerating medium is roughly uniform,
 at least within the timescales we consider.
(Note that this is not a bad approximation for most of the MHD plerion models discussed by KG02.)

The standard expressions
 for the radius $r$ and bulk Lorentz factor $\Gamma$ of the shocked material
 in an adiabatic, spherical blastwave decelerating self-similarly in a uniform medium are
 $r(t) = [12E ct / 4\pi n m_p c^2 (1+z)]^{1/4}
       \sim 3.6 \times 10^{17} {\rm cm} (E_{52}/n)^{1/4} [t_d/(1+z)]^{1/4}$
 and
 $\Gamma(t) = [3E (1+z)^3/ 256\pi n m_p c^5 t^3)]^{1/8}
            \sim 5.9 (E_{52}/n)^{1/8} [t_d/(1+z)]^{-3/8}$,
 respectively (e.g. M\'esz\'aros \& Rees 1997, Piran 1999).
Here $E=10^{52} E_{52} {\rm erg}$ is the blastwave energy,
 $n {\rm cm^{-3}}$ the external medium density,
 $t=t_d {\rm day}$ the observer time elapsed after the GRB,
 $z$ is the GRB redshift,
 and we have adopted the kinematic relation $t={r(1+z) /4\Gamma^2 c}$.
The deceleration starts at radius
 $r_{dec} = (3E / 4\pi n m_p c^2 \Gamma_0^2)^{1/3}
          \sim 2.6 \times 10^{16} {\rm cm} (E_{52}/n)^{1/3} \Gamma_{0,300}^{-2/3}$
 at observer time 
 $t_{dec} = r_{dec}(1+z) /4\Gamma_0^2 c
          \sim 2.4 {\rm s} (1+z) (E_{52}/n)^{1/3} \Gamma_{0,300}^{-8/3}$,
 where $\Gamma_0=300 \Gamma_{0,300}$ is the initial bulk Lorentz factor.

To describe the time-dependent, multiwavelength afterglow spectrum from the blastwave,
 we follow standard discussions of the synchrotron emission
 (e.g. M\'esz\'aros \& Rees 1997, Sari, Piran \& Narayan 1998, Wijers \& Galama 1999),
 and extend it to include cooling and emission by the EC process
 in addition to the synchrotron-self-Compton (SSC) process
 (Panaitescu \& Kumar 2000; Sari \& Esin 2001, hereafter SE01;
  Zhang \& M\'esz\'aros 2001, hereafter ZM01).
It is assumed that constant fractions
 $\epsilon_B=0.01 \epsilon_{B,-2}$ and $\epsilon_e= 0.1 \epsilon_{e,-1}$ of the postshock energy
 are imparted to magnetic field and relativistic electrons, respectively.
The comoving magnetic field is
 $B(t) = (32\pi \epsilon_B n m_p)^{1/2} \Gamma c
       \sim 0.23 {\rm G} \epsilon_{B,-2}^{1/2} E_{52}^{1/8} n^{3/8} [t_d/(1+z)]^{-3/8}$.
Electrons are accelerated in the shock to a power-law distribution
 $dn/d\gamma \propto \gamma^{-p}$
 in the Lorentz factor range $\gamma_m \le \gamma \le \gamma_M$.
We assume $p>2$, our standard choice being $p=2.5$.
The minimum Lorentz factor $\gamma_m$ is given by
 \begin{eqnarray}
 \gamma_m(t) &=& \epsilon_e \Gamma(t) {m_p \over m_e} {p-2 \over p-1} \nonumber\\
             &\sim& 360 \epsilon_{e,-1} [3(p-2)/(p-1)] \nonumber\\
             & & \times (E_{52}/n)^{1/8} [t_d/(1+z)]^{-3/8}
 \label{eqn:gammn}
 \end{eqnarray}
 (note that $[3(p-2)/(p-1)]=1$ when $p=2.5$).
The electrons radiatively cool by the combination of
 the synchrotron, SSC and EC processes,
 the timescales of which are
 $t'_{sy} \sim 6\pi m_e c / \sigma_T B^2 \gamma$,
 $t'_{ssc} \equiv Y t'_{sy}$ and $t'_{ec} \equiv X t'_{sy}$ respectively,
 the total cooling time being
 $t'_c=[(1/t'_{sy})+(1/t'_{ssc})+(1/t'_{ec})]^{-1} = t'_{sy}(1+Y+X)^{-1}$.
The maximum Lorentz factor $\gamma_M$ is determined by balancing $t'_c$ with
 the acceleration time $t'_{acc} \sim 2\pi \gamma m_e c / e B$,
 \begin{eqnarray}
 \gamma_M(t) &=& \left({3e \over \sigma_T B} {1 \over 1+Y+X}\right)^{1/2} \nonumber\\
             &\sim& 0.97 \times 10^8 (1+Y+X)^{-1/2} \epsilon_{B,-2}^{-1/4} \nonumber\\
             & & \times E_{52}^{-1/16} n^{-3/16} [t_d/(1+z)]^{3/16}
 \label{eqn:gammx}
 \end{eqnarray}
 (ZM01).
The cooling Lorentz factor $\gamma_c$ is where $t'_c$ equals
 the comoving adiabatic expansion time $t'_{ad} \sim r/c\Gamma \sim t\Gamma$,
 \begin{eqnarray}
 \gamma_c(t) &=& {1 \over 1+Y+X} {6\pi m_e c \over \sigma_T \Gamma B^2 t} \nonumber\\
             &\sim& 2.9 \times 10^4 (1+Y+X)^{-1} \epsilon_{B,-2}^{-1} \nonumber\\
             & & \times E_{52}^{-3/8} n^{-5/8} [t_d/(1+z)]^{1/8} .
 \label{eqn:gamc}
 \end{eqnarray}
The observed characteristic synchrotron emission frequency
 for an electron of Lorentz factor $\gamma$ is
 $\nu = (4/3) \Gamma (3/4\pi) (eB/m_e c) \gamma^2 (1+z)^{-1}$ (Wijers \& Galama 1999),
 so those for each of $\gamma_m$, $\gamma_M$ and $\gamma_c$
 are respectively
 \begin{eqnarray}
 \nu_m(t) \sim 1.0 \times 10^{12} {\rm Hz}
          \epsilon_{B,-2}^{1/2} \epsilon_{e,-1}^2 [3(p-2)/(p-1)]^2 \nonumber\\
          \times E_{52}^{1/2} t_d^{-3/2} (1+z)^{1/2} ,
 \label{eqn:numn}
 \end{eqnarray}
 \begin{eqnarray}
 \nu_M(t) \sim 0.72 \times 10^{23} {\rm Hz}
          (1+Y+X)^{-1} \nonumber\\
          \times E_{52}^{1/8} n^{-1/8} t_d^{-3/8} (1+z)^{-5/8} ,
 \label{eqn:numx}
 \end{eqnarray}
 and
 \begin{eqnarray}
 \nu_c(t) \sim 6.4 \times 10^{15} {\rm Hz}
          (1+Y+X)^{-2} \epsilon_{B,-2}^{-3/2} \nonumber\\
          \times E_{52}^{-1/2} n^{-1} t_d^{-1/2} (1+z)^{-1/2} .
 \label{eqn:nuc}
 \end{eqnarray}

At early times ($t<t_t$), $\gamma_m > \gamma_c$,
 and all electrons cool within $t'_{ad}$ (fast cooling regime).
The synchrotron flux $f_\nu$ then peaks at $\nu_c$ 
 and mainly consists of three power-law segments:
 $f_\nu=(\nu/\nu_c)^{1/3} f_{\nu,\max}$ for $\nu<\nu_c$,
 $f_\nu=(\nu/\nu_c)^{-1/2} f_{\nu,\max}$ for $\nu_c<\nu<\nu_m$ and
 $f_\nu=(\nu_m/\nu_c)^{-1/2} (\nu/\nu_m)^{-p/2} f_{\nu,\max}$ for $\nu_m<\nu<\nu_M$.
\footnote{
Synchrotron self-absorption
 should cause an additional spectral break
 (or possibly two breaks for fast cooling; Granot, Piran \& Sari 2000)
 at the lowest frequencies,
 but is ignored here
 as it is observationally irrelevant for the high energy afterglow emission,
 our main concern.
}
At late times ($t>t_t$), $\gamma_m < \gamma_c$,
 and only electrons of $\gamma > \gamma_c$ cool within $t'_{ad}$ (slow cooling regime).
Then the spectrum peaks at $\nu_m$, and again has three parts:
 $f_\nu=(\nu/\nu_m)^{1/3} f_{\nu,\max}$ for $\nu<\nu_m$,
 $f_\nu=(\nu/\nu_m)^{-(p-1)/2} f_{\nu,\max}$ for $\nu_m<\nu<\nu_c$ and
 $f_\nu=(\nu_c/\nu_m)^{-(p-1)/2} (\nu/\nu_c)^{-p/2} f_{\nu,\max}$ for $\nu_c<\nu<\nu_M$.
The peak synchrotron flux $f_{\nu,\max}$ for either fast or slow cooling is
 \begin{eqnarray}
 f_{\nu,\max} &=& C \Gamma B r^3 (1+z)/D_L^2 \nonumber\\
    &\sim& 2.9 {\rm mJy} \epsilon_{B,-2}^{1/2} E_{52} n^{1/2} D_{L,28}^{-2} (1+z) ,
 \label{eqn:fmax}
 \end{eqnarray}
 where $C$ is a numerical constant and $D_L=10^{28} D_{L,28} {\rm cm}$
 is the luminosity distance to the GRB.
The transition from fast to slow cooling ($\gamma_c=\gamma_m$) occurs at the time
 \begin{eqnarray}
 t_t
     \sim 13 {\rm s} (1+Y+X)^2 \epsilon_{B,-2}^2 \epsilon_{e,-1}^2 [3(p-2)/(p-1)]^2 \nonumber\\
     \times E_{52} n (1+z) .
 \label{eqn:tt}
 \end{eqnarray}

To account for the effects of EC and SSC cooling,
 we have introduced $X$ and $Y$, which are in general time-dependent.
The EC parameter $X$ equals the ratio of comoving frame energy density
 in plerion radiation field $u'_p$ to that in magnetic field $u'_B$.
If the blastwave decelerates inside the plerion ($r<R_p$),
 which is the case mainly considered below,
 we may presume that the radiation field is nearly uniform and isotropic, obtaining
 \begin{eqnarray}
 X &\sim& {u'_p \over u'_B}
   \sim \Gamma^2 {\xi_e L_p \over 4\pi R_p^2 c} \nonumber\\
   & & \times {\ln(\min\{\nu'^p_M,\nu'^p_{KN}\}/\nu'^p_a) \over 2 \ln(\gamma^p_M/\gamma^p_m)}
         /4 \epsilon_B \Gamma^2 n m_p c^2 ,
 \label{eqn:xpar}
 \end{eqnarray}
 where $\nu'^p_{KN} \sim m_e c^2/h \gamma$
 is the frequency above which Klein-Nishina (KN) effects suppress Compton scattering,
 depending on $\gamma$.
Observe here that the factors of $\Gamma^2$ in $u'_p$ and $u'_B$ cancel out
 so that aside from the weak $t$-dependence through $\nu'^p_{KN}$, $X$ is constant in time.
Generally $\nu'^p_M > \nu'^p_{KN}$,
 but we do not make large errors in taking $\min\{\nu'^p_M,\nu'^p_{KN}\} \sim \nu'^p_M$
 in the logarithmic factor for simplicity.
For a blastwave decelerating outside the plerion ($r>R_p$),
 the radiation field is anisotropic and falls with $r$,
 and $X$ differs from eq.\ref{eqn:xpar}
 by a factor of $\sim (R_p/r)^2 (1-2\mu+4\mu^2)/3$ where $\mu=\sqrt{1-(R_p/r)^2}$
 (which applies for $\Gamma \gg 1$; see e.g. Sikora et al. 1996, Inoue \& Takahara 1997),
 decreasing with time.

The SSC parameter $Y$ can be evaluated as in SE01,
 $Y \sim u'_{sy}/u'_B \sim \eta u'_e/[u'_B (1+Y+X)] \sim \eta \epsilon_e/[\epsilon_B (1+Y+X)]$,
 where $u'_e$ and $u'_{sy}$ are the comoving frame energy densities
 in relativistic electrons and synchrotron radiation,
 and $\eta \sim \min\{1,(\gamma_c/\gamma_m)^{2-p}\}$ is the fractional energy radiated by electrons.
In the fast cooling phase, $\eta \sim 1$ and we get the expression
 \begin{eqnarray}
 Y \sim {1 \over 2}\left(\sqrt{4{\epsilon_e \over \epsilon_B}+(1+X)^2}-1-X\right) .
 \label{eqn:yf}
 \end{eqnarray}
In the slow cooling phase,
 $Y$ decreases with time but only very slowly for typical values of $p$ (SE01),
 so we simplify by assuming eq.\ref{eqn:yf} at all times.

The SSC spectrum has a similar shape to the synchrotron spectrum,
 and we approximate it as comprising broken power-laws (SE01, ZM01),
 with characteristic frequencies at
 \begin{eqnarray}
 \nu_m^{SC} &\sim& \gamma_m^2 \nu_m \nonumber\\
    &\sim& 1.2 \times 10^{17} {\rm Hz} \epsilon_{e,-1}^4 \epsilon_{B,-2}^{1/2} [3(p-2)/(p-1)]^4 \nonumber\\
                                       & & \times  E_{52}^{3/4} n^{-1/4} t_d^{-9/4} (1+z)^{5/4}
 \label{eqn:numinsc}
 \end{eqnarray}
 and
 \begin{eqnarray}
 \nu_c^{SC} &\sim& \gamma_c^2 \nu_c \nonumber\\
    &\sim& 5.4 \times 10^{24} {\rm Hz} (1+Y+X)^{-4} \epsilon_{B,-2}^{-7/2} \nonumber\\
               & & \times E_{52}^{-5/4} n^{-9/4} t_d^{-1/4} (1+z)^{-3/4} .
 \end{eqnarray}
The maximum SSC frequency should occur at
 \begin{eqnarray}
 \nu_{KN} &\sim& \Gamma \gamma_M m_e c^2 (1+z)^{-1} \nonumber\\
          &\sim& 0.71 \times 10^{28} {\rm Hz} (1+Y+X)^{-1/2} \epsilon_{B,-2}^{-1/4} \nonumber\\
                      & & \times E_{52}^{1/16} n^{-5/16} t_d^{-3/16} (1+z)^{-13/16} ,
 \label{eqn:nukn}
 \end{eqnarray}
 where KN effects completely cut off the spectrum.
(Generally, $\nu_{KN} \ll \nu_M^{SC} \sim \gamma_M^2 \nu_M$.)
For fast cooling, the SSC flux $f_\nu^{SC}$ peaks at $\nu_c^{SC}$,
 the spectrum being $\propto \nu^{1/3}$ for $\nu<\nu_c^{SC}$,
 $\propto \nu^{-1/2}$ for $\nu_c^{SC}<\nu<\nu_m^{SC}$
 and $\propto \nu^{-p/2}$ for $\nu_m^{SC}<\nu<\nu_{KN}$;
 for slow cooling, the peak is at $\nu_m^{SC}$,
 and the spectrum is $\propto \nu^{1/3}$ for $\nu<\nu_m^{SC}$,
 $\propto \nu^{-(p-1)/2}$ for $\nu_m^{SC}<\nu<\nu_c^{SC}$
 and $\propto \nu^{-p/2}$ for $\nu_c^{SC}<\nu<\nu_{KN}$.
The peak SSC flux $f_{\nu,\max}^{SC}$
 is equal to the peak synchrotron $f_{\nu,\max}$
 multiplied by the optical depth $\tau_e$ of the shocked material,
 \begin{eqnarray}
 f_{\nu,\max}^{SC} &\sim& \tau_e f_{\nu,\max} \sim {\sigma_T n r \over 3} f_{\nu,\max} \nonumber\\
  &\sim& 2.3 \times 10^{-6} {\rm mJy} \epsilon_{B,-2}^{1/2} E_{52}^{5/4} n^{5/4} D_{L,28}^{-2} t_d^{1/4} (1+z)^{3/4} .
 \label{eqn:fmaxsc}
 \end{eqnarray}

The EC spectrum reflects the shapes of
 both the electron distribution and the plerion spectrum.
Considering deceleration inside the plerion,
 and again adopting the broken power-law approximation,
 spectral breaks arise at the frequencies where electrons of $\gamma_m$ and $\gamma_c$
 upscatter plerion photons of $\nu^p_a$,
 \begin{eqnarray}
 \nu_m^{EC} &\sim& \gamma_m^2 \Gamma^2 \nu^p_a (1+z)^{-1} \nonumber\\
            &\sim& 1.1 \times 10^{18} {\rm Hz} \epsilon_{e,-1}^2 [3(p-2)/(p-1)]^2 \nonumber\\
                                               & & \times E_{52}^{1/2} n^{-1/2} t_d^{-3/2} (1+z)^{1/2}
 \label{eqn:numinec}
 \end{eqnarray}
 and
 \begin{eqnarray}
 \nu_c^{EC} &\sim& \gamma_c^2 \Gamma^2 \nu^p_a (1+z)^{-1} \nonumber\\
            &\sim& 0.71 \times 10^{22} {\rm Hz} (1+Y+X)^{-2} \epsilon_{B,-2}^{-2} \nonumber\\
                                              & & \times E_{52}^{-1/2} n^{-3/2} t_d^{-1/2} (1+z)^{-1/2} ,
 \end{eqnarray}
 where the fiducial numerical value for $\nu^p_a$ in eq.\ref{eqn:nupa} has been used.
Note the factor $\Gamma^2$ which results from two Lorentz transformations,
 one into the comoving frame from the observer frame and another vice-versa.
The fast cooling EC spectrum has a peak flux at $\nu_c^{EC}$,
 is $\propto \nu$ for $\nu<\nu_c^{EC}$
 from the behavior of the Compton scattering cross section,
 $\propto \nu^{-1/2}$ for $\nu_c^{EC}<\nu<\nu_m^{EC}$
 and $\propto \nu^{-1}$ for $\nu_m^{EC}<\nu<\nu_{KN}$,
 mirroring the flat plerion spectrum.
The KN limit discussed above also applies here for the maximum EC frequency,
 $\nu^p_M$ being irrelevant.
Likewise, the slow cooling EC spectrum peaks at $\nu_m^{EC}$,
 and is $\propto \nu$ for $\nu<\nu_m^{EC}$,
 $\propto \nu^{-(p-1)/2}$ for $\nu_m^{EC}<\nu<\nu_c^{EC}$
 and $\propto \nu^{-1}$ for $\nu_c^{EC}<\nu<\nu_{KN}$.
The peak EC flux $f_{\nu,\max}^{EC}$ may be obtained analogously to the SSC case,
 but using the relation
 $f'^{EC}_{\nu',\max} \sim \tau_e f'^p_{\nu'^p,\max}$ in the comoving frame,
 where $f'^p_{\nu'^p,\max} \sim \Gamma f^p_{\nu^p,\max}$
 is the the peak plerion flux emitted at $\nu'^p=\nu'^p_a \sim \Gamma \nu^p_a$.
Accounting for Lorentz transformations to and from the comoving frame,
 \begin{eqnarray}
 f_{\nu,\max}^{EC} &\sim& \Gamma^2 \tau_e f^p_{\nu,\max} {r^2 \over D_L^2} (1+z) \nonumber\\
    &=& \Gamma^2 {\sigma_T n r \over 3}
      {\xi_e L_p (\nu^p_a)^{-1} \over 8\pi R_p^2 \ln(\gamma^p_M/\gamma^p_m)} {r^2 \over D_L^2} (1+z) \nonumber\\
    &\sim& 2.3 \times 10^{-3} {\rm mJy} E_{52} D_{L,28}^{-2} (1+z) ,
 \label{eqn:fec}
 \end{eqnarray}
 where the last line assumes fiducial plerion parameters.
We caution that being isotropic in the observer frame, the plerion radiation field
 should actually be highly anisotropic in the shock frame moving at $\Gamma \gg 1$.
Instead of an accurate but cumbersome calculation
 of the Compton upscattered photon distribution
 including the full angle-dependence,
 we have approximated by using quantities averaged over angles in the comoving frame;
 see e.g. Dermer 1995, Inoue \& Takahara 1996, Sikora 1997.

The broken power-law representations of the EC and SSC spectra
 should be adequate to within an order of magnitude,
 but more accurate calculations properly integrating over the broad seed photon frequency distribution
 will result in smoother spectral breaks
 and somewhat larger fluxes (above $\nu_c^{EC}$ and $\nu_c^{SC}$ for fast cooling,
 or above $\nu_m^{EC}$ and $\nu_m^{SC}$ for slow cooling)
 due to logarithmic terms contributed by a range of electron Lorentz factors (SE01).
Regarding KN effects,
 a gradual steepening from $\nu \lesssim \nu_{KN}$
 instead of an abrupt cutoff should actually occur,
 as parts of the seed photon spectra will be in the KN regime 
 even for electrons with $\gamma \lesssim \gamma_M$.
We have not included the effects of internal pair attenuation,
 which could be important for the highest emission energies
 at relatively early times in the afterglow.
However, significant differences are expected only for TeV energies and above (ZM01),
 whereby pair degradation by the infrared background during intergalactic propagation 
 should be serious anyway for GRBs with $z \gtrsim 0.1$
 (e.g. Stecker \& de Jager 1998, Totani 2000).

\subsection{Constraints on the Afterglow Environment}
\label{sec:aftenv}

Having laid out the tools to calculate the broadband afterglow emission,
 we now discuss some constraints
 to be imposed on the local environment of the blastwave
 in the context of our model.
For an afterglow to produce a strong Fe line as observed in GRB991216,
 it is necessary that the GRB blastwave decelerates
 inside the plerion near relatively dense SNR fragments,
 so that they are irradiated efficiently by afterglow continuum X-rays.

In case the decelerating medium comprises mostly entrained baryons
 (referred to as the ``baryon case''),
 we opt to directly use the above equations
 with $\epsilon_{B,-2} = 1$ and $\epsilon_{e,-1} = 1$,
 since the physics governing these quantities here
 should be similar to that for conventional afterglows decelerating in external media.
Although an inhomogeneous spatial distribution is actually expected for the baryons,
 we surmise that the formulae are applicable for a certain timescale
 during which the medium can be approximated as locally uniform.
The local baryon density in the plerion interior
 may range anywhere from values as high as $n \sim 10^{11} {\rm cm^{-3}}$
 for the densest SNR clumps,
 to much lower values $n \ll 1 {\rm cm^{-3}}$
 for regions efficiently swept out by the magnetized plerionic plasma.
If the local density in our line of sight
 $n \gtrsim 10^6 {\rm cm^{-3}}$,
 the blastwave turns subrelativistic
 in less than $\sim 1 {\rm day}$
 and is unable to generate the observed afterglows
 (Lazzati et al. 1999; see however Masetti et al. 2001, In 't Zand et al. 2001),
 whereas if the density $n \lesssim 10^{-3} {\rm cm^{-3}}$,
 the blastwave does not decelerate significantly
 until it reaches the medium outside the plerion at $r > R_p$.
We fiducially choose $n \sim 10^3 {\rm cm^{-3}}$,
 implying that if the medium was globally uniform at this value,
 the blastwave travels
 from $r \sim 3 \times 10^{15} {\rm cm}$ to $\sim 6 \times 10^{16} {\rm cm}$ 
 during $t \sim 0.2$ s to $\sim 2$ days inside the plerion,
 illuminating nearby Fe-rich condensations along the way,
 and turns subrelativistic ($\Gamma \sim 1$)
 at $t \sim 15$ days.
In reality, a sufficiently clumpy baryonic distribution
 may cause a sporadic type of variability in the afterglow
 which should be detectable by future instruments,
 and could perhaps be relevant to
 the X-ray flaring behavior observed in the afterglows of
 GRB970508 (Piro et al. 1999) and GRB970828 (Yoshida et al. 2001).

For the case of deceleration occurring
 mainly within the magnetized plerionic material
 (referred to as the ``plerion case''),
 KG02 showed that the usual afterglow formulae can be utilized with some modifications,
 such as substituting $n$ with an equivalent baryon density
 \begin{eqnarray}
 n_{b,eq}={1 \over m_p c^2} \left(4p_e+{(B'+E')^2 \over 2\pi}\right) ,
 \label{eqn:nbeq}
 \end{eqnarray}
 where $p_e$ is the electron pressure
 and $B'$ and $E'$ are the magnetic and electric fields
 in the frame comoving with the shocked pulsar wind.
For our simplified plerion model,
 we estimate $n_{b,eq}$ by taking
 $E'=0$, $B'=B_p=(3 \xi_B E_p / R_p^3)^{1/2}$ and $p_e=\xi_e E_p / 8\pi R_p^3$,
 which gives
 $n_{b,eq} \simeq 10^3 {\rm cm^{-3}} (2\xi_{e,0.5}/3+\xi_{B,0.5}) \ t_{p,120}^{-3}$.
Thus, as with the baryon case, we fiducially choose $n=10^3 {\rm cm^{-3}}$
 for the plerion case.
Spatial uniformity should be a good approximation,
 even for the more sophisticated MHD models considered by KG02.
One must also make corresponding changes for $\epsilon_e$ and $\epsilon_B$,
 as well as insert correction factors for some quantities,
 but these depend on the detailed properties of the pulsar wind shock (see KG02)
 for which our model lacks information.
For want of a better alternative,
 we choose to employ $\epsilon_e = \xi_e =0.5$ and  $\epsilon_B = \xi_B =0.5$,
 and ignore the correction factors.
With these parameters,
 the blastwave should be partially radiative during the fast cooling stage;
 however, an accurate treatment of such cases is rather cumbersome
 (e.g. B\"ottcher \& Dermer 2000),
 so we proceed with the adiabatic assumption for the sake of simplicity.

A further requirement from the observed Fe lines is that
 $t_p$ should be neither much shorter nor much longer than $\sim 120 {\rm days}$:
 $t_{p,120} \ll 1$ would not allow time
 for enough SN-synthesized radioactive Co to decay to Fe (Vietri et al. 2001),
 whereas for $t_{p,120} \gg 1$ the SNR fragments would be of too low density
 to generate the observed Fe line flux,
 unless they can be made extremely condensed (eqs.\ref{eqn:nl},\ref{eqn:nes}).
This justifies our fiducial choice of $t_{p,120}=1$.
\footnote{
These conditions may be consistent
 with the time-delay and ionization parameter constraints
 discussed e.g. in Piro et al. (2000) and Vietri et al. (2001),
 since they apply to the distance $d_i$ between the continuum source
 and the irradiated, line-emitting matter, which can be smaller than
 the radius $r$ of the matter from the GRB site.
The blastwave (i.e. the continuum source) itself
 should have moved out to $r \gtrsim 10^{16}-10^{17} {\rm cm}$ by $t \sim 1 {\rm day}$,
 but Fe-rich material could be present in its immediate vicinity at $d_i \ll r$.
}

\subsection{Results and Discussion}
\label{sec:res}

Guided by the observed characteristics of afterglows such as GRB991216,
 our fiducial parameters were chosen to represent
 conditions favorable for generating strong Fe lines.
However,
 we will not attempt model fits
 to the available data on individual objects,
 since i) realistically, additional complicating effects of jet geometry, etc.
 can be important and must be taken into account,
 and ii) full parameter searches for fitting particular observations are out of our scope
 (c.f. Panaitescu \& Kumar 2002)
The results given here for the fiducial parameter sets
 are intended to exemplify some notable features of afterglows
 wherein EC cooling and emission induced by the supranova plerion are important.
Changing these parameters can lead to either stronger or weaker EC effects (see below).

\begin{figure}
 \figurenum{2} 
 \epsscale{0.6} 
 \plotone{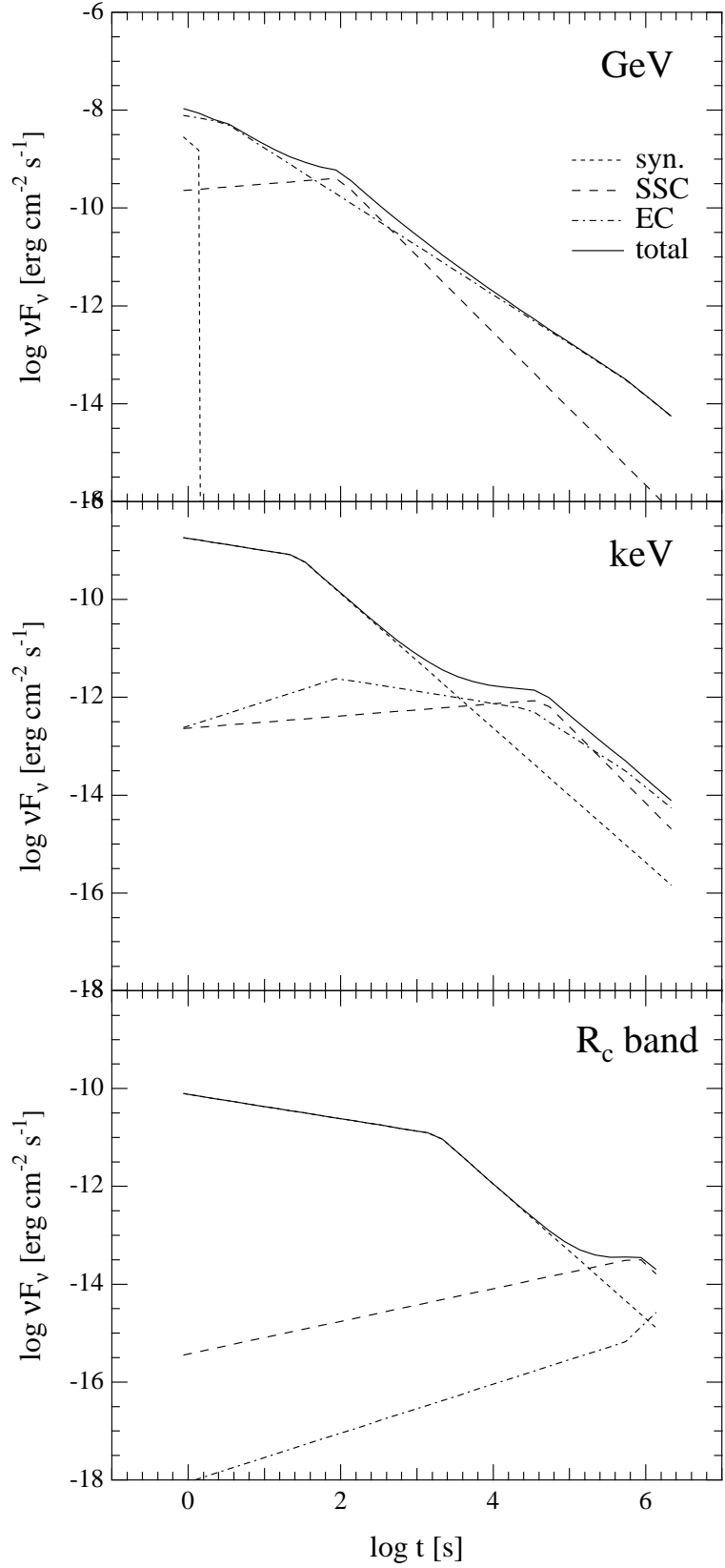} 
 \caption{
The afterglow light curves for the fiducial baryon case
 at $\nu=2.4 \times 10^{23} {\rm Hz}$ (1 GeV, top),
 $\nu=2.4 \times 10^{17} {\rm Hz}$ (1 keV, middle) and
 $\nu=4.5 \times 10^{14} {\rm Hz}$ ($R_c$ band, bottom).
The synchrotron (short-dashed), SSC (long-dashed) and EC (dot-dashed) components
 are also shown separately.
} 
 \label{fig:lcb}
 \end{figure}

We first discuss the baryon case.
Figs.\ref{fig:specb} and \ref{fig:lcb}
 show our fiducial results in terms of
 the time-dependent broadband afterglow spectra at different observer times $t$,
 and the afterglow light curves at selected fixed frequencies, respectively,
 assuming the GRB to be at $z=1$ in the flat lambda cosmology
 (i.e. $D \sim 2 \times 10^{28} {\rm cm}$).

The values of $X$ and $Y$ here are fiducially $X \sim 1.3$ and $Y \sim 2.2$,
 and since they were approximated as being constant,
 the relative luminosities of the synchrotron, SSC and EC components stay roughly the same.
However, each component is prominent in a different energy band which changes with $t$:
 at the beginning of the afterglow ($t \lesssim 1 {\rm min}$),
 the SSC component dominates at the highest energies ($\sim$ TeV)
 whereas the EC component does so at somewhat lower energies ($\sim$ GeV);
 in the latter stages ($t \gtrsim 1 {\rm hr}$), this is reversed.
The lower energies are always dominated by the synchrotron component,
 although the energy of the luminosity peak progressively decreases
 from $\sim$ MeV down to the radio band.
There are clear differences between the spectral shapes of the SSC and EC components:
 the SSC peaks in $\nu f_{\nu}$ at $\nu^{SC} \sim \max\{\nu_m^{SC},\nu_c^{SC}\}$
 and has a steep high energy slope above this break
 identical to that of the synchrotron emission (energy index $-p/2$),
 but the EC component is flat in $\nu f_{\nu}$ above $\nu^{EC} \sim \max\{\nu_m^{EC},\nu_c^{EC}\}$,
 reflecting the harder plerion spectrum.
This spectral signature is one way to observationally distinguish between SSC and EC emission,
 even though this distinction can be blurred if values of $p$ happen to be close to $s$.

Here the fast to slow cooling transition occurs at $t_t \sim 6.3 {\rm days}$,
 and the afterglow is still fast cooling at $t \sim 1 {\rm day}$.
This is longer than in standard discussions,
 and is partly due to the extra cooling process of EC,
 but primarily as a consequence of the relatively high external density adopted,
 $n \sim 10^3 {\rm cm^{-3}}$.
The light curves at fixed frenquencies can be described by breaks at
 characteristic times $t_m$ and $t_c$
 corresponding to the passage of the minimum and cooling frequencies for each spectral component,
 as well as at $t_t$.
The synchrotron light curves are as in previous studies (Sari et al. 1998):
 for high frequencies in which $t_c < t_m < t_t$,
 $f_{\nu} \propto t^{1/6}$ for $t < t_c$,
 $f_{\nu} \propto t^{-1/4}$ for $t_c < t < t_m$ and
 $f_{\nu} \propto t^{(2-3p)/4}$ [$t^{-1.38}$] for $t > t_m$;
 for low frequencies in which $t_t < t_m < t_c$,
 $f_{\nu} \propto t^{1/6}$ for $t < t_t$,
 $f_{\nu} \propto t^{1/2}$ for $t_t < t < t_m$,
 $f_{\nu} \propto t^{(3-3p)/4}$ [$t^{-1.13}$] for $t_m < t < t_c$ and
 $f_{\nu} \propto t^{(2-3p)/4}$ [$t^{-1.38}$] for $t > t_c$.
In these expressions and below, the $t$-dependence inside the brackets
 are for our standard choice of $s=2$ and $p=2.5$.
Defining $t_c^{SC}$, $t_m^{SC}$ as times when $\nu=\nu_c^{SC}$, $\nu_m^{SC}$ respectively,
 the SSC light curves are:
 for high frequencies in which $t_c^{SC} < t_m^{SC} < t_t$,
 \begin{eqnarray}
 f_{\nu}^{SC} \propto
 \left\{
 \begin{array}{ll}
    t^{1/3}  &  t < t_c^{SC}\\
    t^{1/8}  &  t_c^{SC} < t < t_m^{SC}\\
    t^{(10-9p)/8} \ \ [t^{-1.56}]  &  t > t_m^{SC} ;
 \end{array}
 \right.
 \label{eqn:fhsc}
 \end{eqnarray}
 for low frequencies in which $t_t < t_m^{SC} < t_c^{SC}$,
 \begin{eqnarray}
 f_{\nu}^{SC} \propto
 \left\{
 \begin{array}{ll}
    t^{1/3}  &  t < t_t\\
    t  &  t_t < t < t_m^{SC}\\
    t^{(11-9p)/8} \ \ [t^{-1.44}]  &  t_m^{SC} < t < t_c^{SC}\\
    t^{(10-9p)/8} \ \ [t^{-1.56}]  &  t > t_c^{SC} .
 \end{array}
 \right.
 \label{eqn:flsc}
 \end{eqnarray}
The EC light curves depend on the plerion spectral index ($-s/2$).
With $t_c^{EC}$, $t_m^{EC}$ as times when $\nu=\nu_c^{EC}$, $\nu_m^{EC}$ respectively:
 for high frequencies in which $t_c^{EC} < t_m^{EC} < t_t$,
 \begin{eqnarray}
 f_{\nu}^{EC} \propto
 \left\{
 \begin{array}{ll}
    t^{1/2}  &  t < t_c^{EC}\\
    t^{-1/4}  &  t_c^{EC} t < t_m^{EC}\\
    t^{(2-3s)/4} \ \ [t^{-1}]  &  t_m^{EC} < t < t_t\\
    t^{(2-2p-s)/4} \ \ [t^{-1.2}]  &  t > t_t ;
 \end{array}
 \right.
 \label{eqn:fhec}
 \end{eqnarray}
 for low frequencies in which $t_t < t_m^{EC} < t_c^{EC}$,
 \begin{eqnarray}
 f_{\nu}^{EC} \propto
 \left\{
 \begin{array}{ll}
    t^{1/2}  &  t < t_t\\
    t^{3/2}  &  t_t < t < t_m^{EC}\\
    t^{(3-3p)/4} \ \ [t^{-1.13}]  &  t_m^{EC} < t < t_c^{EC}\\
    t^{(2-2p-s)/4} \ \ [t^{-1.2}]  &  t > t_c^{EC} .
 \end{array}
 \right.
 \label{eqn:flec}
 \end{eqnarray}

Differences can be seen in the high frequency decay indices at late times
 between the SSC and EC components:
 $f_{\nu}^{SC} \propto t^{(10-9p)/8}$ [$t^{-1.56}$] both before and after $t_t$ for SSC,
 compared to $f_{\nu}^{EC} \propto t^{(2-3s)/4}$ [$t^{-1}$] before $t_t$
 breaking to $f_{\nu}^{EC} \propto t^{(2-2p-s)/4}$ [$t^{-1.2}$] after $t_t$ for EC.
These features should be important observational diagnostics in the X-ray and gamma-ray range,
 although here again, the discrimination is less clear if $p$ approaches $s$.
(Note also that there should actually be a slight break in the SSC light curve at $t_t$
 from the time-dependence of $Y$; SE01.)
The generally harder spectrum of the EC emission relative to SSC should make it
 increasingly dominant in the higher energy bands at later times.

The early X-ray and optical emission is predominantly synchrotron,
 but the entry of the minimum frequency of the SSC emission
 (and to a lesser extent, the EC emission)
 gives rise to `bumps' in the light curves,
 which appear at $t \sim 0.5$ days in X-rays and $t \sim 10$ days in the optical
 for our fiducial calculation.
In particular, we see that X-ray emission after $\sim 1$ day is SSC plus EC emission.
So far, there have been no clear indications for such X-ray bumps,
 except for a flattening in the X-ray light curve of GRB000926 at $t \sim 2$ days.
Together with multiwavelength spectra,
 this has been interpreted as SSC emission from a blastwave in a moderately dense medium
 of $n \sim 30 {\rm cm^{-3}}$ (Harrison et al. 2001),
 but a different explanation with an even denser medium ($n \sim 4 \times 10^4 {\rm cm^{-3}}$)
 is also possible (Piro et al. 2001).
In the optical band, it is interesting to recall that a small number of afterglows
 have been seen to exhibit bumps at few tens of days,
 attributable to an underlying supernova light curve
 superimposed on the afterglow
 (e.g. Bloom et al. 1999, 2002, Galama et al. 2000, Lazzati et al. 2001b.
The observed bumps seem to manifest simultaneous spectral reddening,
 as opposed to the spectral hardening expected for the SSC induced bumps.
However, with the exception of GRB980326 and GRB011121,
 the combined evidence for light curve bump plus reddening is not very strong,
 so an SSC explanation may not be ruled out, at least for some of the observations.
In any case, realistic modeling of light curve breaks
 should also include other effects such as jet geometry,
 so firm conclusions in comparison with observations are not possible at the moment.

The detection of the EC component, which is distinctive to the supranova model,
 should be best achieved at GeV energies and above.
We can assess the detectability in GeV gamma-rays by the EGRET and GLAST instruments,
 following the discussion of ZM01.
The fluence threshold of each intrument are described
 by a constant value for photon-limited, short integration times  
 and one proportional to $t^{1/2}$ for background-limited, long integration times.
This can be compared with the model fluence light curves,
 intergrated over the energy range of 400 MeV to 200 GeV.
Our fiducial baryon case model would have been observable by EGRET
 hours after the GRB from $z \lesssim 0.5$,
 providing a viable explanation for the extended GeV emission seen in GRB940217 (Hurley et al. 1994).
Similar observations should also be possible with AGILE (e.g. Mereghetti et al. 2001),
 having a slightly better sensitivity than EGRET.
GLAST should be able to detect this emission
 from redshifts as high as $z \sim 1.5$ up to $\sim$ 1 day after the burst.

\begin{figure}
 \figurenum{4} 
 \epsscale{0.6}
 \plotone{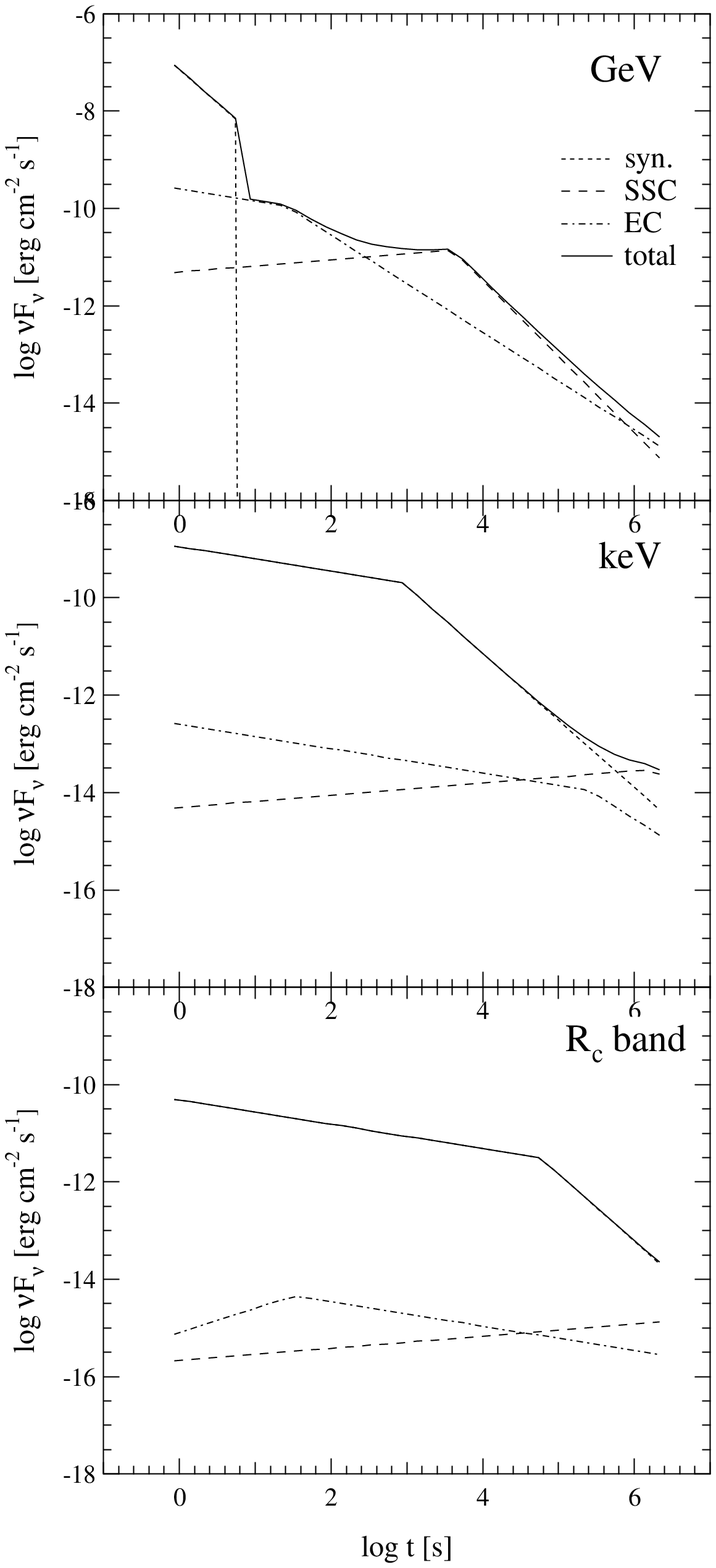} 
 \caption{
As with fig.\ref{fig:lcb}, but for the fiducial plerion case.
} 
 \label{fig:lcp}
 \end{figure}

Our results for the fiducial plerion case are shown in Figs.\ref{fig:specp} and \ref{fig:lcp}.
Since the equivalent baryon density $n_{b,eq}=10^3 {\rm cm^{-3}}$,
 the only difference here with the fiducial baryon case
 stems from the parameter values of $\epsilon_e$ and $\epsilon_B$.
Many of the trends described above are also valid here qualitatively,
 but important quantitative differences are to be seen.
The much larger magnetic field (by a factor of 50 in energy density)
 induces greatly increased synchrotron losses,
 leading to smaller values of $\gamma_c$ and the corresponding cooling frequencies,
 while the larger electron injection efficiency (by a factor of 5)
 causes higher values of $\gamma_m$ and the associated miminum frequencies.
This implies very large $t_t$, 
 and the electrons remain fast cooling throughout the duration of the afterglow.
The synchrotron and SSC luminosities are larger and occupy higher energy bands;
 the EC emission is relatively much less prominent,
 appearing at GeV over the SSC emission
 only during the early time interval from $\sim 10$ s to $\sim 300$ s,
 and again at later times $\gtrsim 10$ days.
Detection of this component should be feasible with future gamma-ray instruments
 as with the baryon case,
 but here it may be more difficult to entangle from the SSC emission.
The light curve breaks due to the passage of the minimum frequecies occur much later,
 and the SSC bumps are not observable.

We now discuss the effect of changing our main free parameters,
 particularly those regarding the plerionic enviroment, $n$ and $t_p$.
Keeping all others constant for the baryon case,
 decreasing $n$ from $n \sim 10^3 {\rm cm^{-3}}$ 
 leads to a greater dominance of the EC component and less synchrotron emission,
 as this implies a smaller $B$ and larger $X$ for the same $\epsilon_B$.
Higher values of $n$ may also occur inside the plerion;
 note the recent observations of GRB010222 indicating $n \sim 10^6 {\rm cm^{-3}}$
 for this burst (Masetti et al. 2001, In 't Zand et al. 2001),
 although alternative interpretations with lower values of $n$ are possible
 (e.g. Panaitescu \& Kumar 2002).
More `conventional' parameters (lower $n$, small $X$)
 may be accomodated in the context of our model
 by taking a larger $t_p$ to make $u_p \propto L_p/R_p^2 \propto t_p^{-3}$ smaller,
 say $t_p \sim 3$yr, and assuming $n \sim 1 {\rm cm^{-1}}$.
For the plerion case, $n$ is not an independent free parameter as $n=n_{n,eq} \propto t_p^{-3}$;
 here $t_p \sim 3$yr directly implies $n \sim 1 {\rm cm^{-1}}$.
However, for this larger plerion
 the volume-averaged SN ejecta density should also drop (eq.\ref{eqn:nes}),
 so that strong Fe lines are less likely, requiring extreme clumping of ejecta material.
Note that clear upper limits on Fe line emission
 obtained for some bright afterglows
 point to a variety of GRB environments (Yonetoku et al. 2000, 2001),
 which could correspond to a range of $t_p$ and/or $n$ in our model.
We also mention that
 the well observed afterglow of GRB970508
 has an Fe line feature detection (Piro et al. 1999),
 as well as arguments for consistency with more or less `standard' parameters (Wijers \& Galama 1999),
 but including the effects of SSC and EC cooling in the model fitting
 may allow different sets of parameters (e.g. SE01).

\section{Direct Detectability of the Precursor Plerion}
\label{sec:dir}

Finally, we briefly remark on the possibility of the direct detection and identification
 of the precursor plerion emission,
 which would be a straightforward test for the supranova model (VS98).
A more detailed assessment can be found in Guetta \& Granot (2002).

As discussed above,
 the plerion should be a luminous, steady source 
 during a period $t_p$ prior to the burst,
 typically with bolometric luminosity $\sim \xi_e L_p \sim 10^{46} {\rm erg \ s^{-1}} t_{p,120}^{-1}$
 and power-law spectrum of energy index $\sim -1$
 between frequencies $\sim 10^{11} {\rm Hz}$ and $\sim 10^{22} {\rm Hz}$.
This is comparable to high-luminosity quasars and readily observable;
 e.g. it should have already been detected out to $z \sim 0.7$
 as bright keV X-ray sources
 during the ROSAT All-Sky Survey (RASS; Voges et al. 1999)
The event rate of GRBs detectable by BATSE from $z \lesssim 0.7$
 may be roughly
 $R_{GRB} \sim 30 {\rm yr^{-1}}$ all-sky
 (e.g. Porciani \& Madau 2001),
 so there should always be at least
 $R_{GRB} t_p/f_b \sim 10 f_b^{-1}$
 active plerions all-sky above the RASS sensitivity.
Here $f_b^{-1}$ is inverse of the fractional solid angle subtended by the GRB beam,
 which could be as large as $\sim 100$ (Frail et al. 2001),
 and we have assumed that the plerions are not strongly beamed.
However, the problem lies in ascertaining these sources as precursors to GRBs,
 which must be shown to be positionally coincident
 as well as correlated in time within $t_p$
 before the GRB.
Even though the operation periods of BATSE and RASS were close enough in time,
 the large BATSE error boxes ($\Delta\theta \sim 1^{\circ}$) preclude
 discrimination from the large number of AGNs with similar X-ray spectra.
More precise GRB localizations by present and next generation instruments
 (e.g. HETE-2 with $\Delta\theta \sim 10''-10'$ or Swift with $\Delta\theta \sim 1''-4'$),
 combined with concurrently conducted wide field X-ray surveys
 (e.g. XMM-Newton serendipitous survey)
 should improve the prospects on searching for GRB - precursor plerion associations.

The plerion should also be conspicuous in the optical,
 e.g. its R-band magnitude would be $R_c \sim 17$ at $z \sim 1$.
(The precursor supernova may be observable as well, but could be masked by the luminous plerion.)
Again, identifying these with the ensuing GRB out of the numerous other objects
 at similar magnitudes
 requires accurate GRB positions and nearly contemporaneous wide field optical surveys.
This may be possible through serendipitous studies of observations made by
 e.g. the Subaru Suprime-Cam, the Sloan Digital Sky Survey,
 and the future Supernova Acceleration Probe mission.
Detection at radio frequencies may also be possible
 if there is additional emission from the plerion forward shock (\S \ref{sec:ple}).

\section{Summary and Comments}
\label{sec:sum}

We summarize the salient points of our work.
In the supranova model of GRBs,
 the fast-rotating SMNS active during the time between the SN and the GRB
 should drive a luminous plerionic nebula into the preburst environment,
 with a number of important consequences for the ensuing GRB afterglow.
Rayleigh-Taylor instabilities acting at the plerion-SN ejecta interface
 may induce significant filamentary clumping of the ejecta material,
 allowing a favorable geometry and enhanced emissivity for the afterglow Fe line emission.
The plerion radiation field
 can act as seeds for EC cooling and emission in the GRB external shock,
 leading to prominent GeV-TeV gamma-rays during the afterglow,
 which is detectable by GLAST out to typical GRB redshifts,
 and distinguishable from SSC emission by its characteristic spectrum and light curve.
A direct search for the plerion emission prior to the GRB
 may be conducted through accurate GRB positions
 and concurrent wide field surveys, e.g. in the optical and X-rays.
All of these should provide critical tests for the supranova model in the near future.

The ability of the plerion to effectively fragment the SNR
 and penetrate between the ejecta material is an uncertain aspect of our scenario.
We may therefore consider the consequences of the plerion
 being completely confined inside the SNR shell instead.
In the case of a spherical geometry,
 such a circumstance would be at odds
 with constraints derived from the observed Fe lines, since:
 1) the SNR would likely be optically thick to Thomson scattering
 at the time of the GRB, so that no prompt or afterglow emission
 can be seen (eq.\ref{eqn:nes});
 2) the SNR shell, as well as the plerionic bubble material,
 would be dense enough to decelerate the GRB blastwave in a time
 shorter than the observed afterglow; and
 3) the radiation density of the plerion would be so large
 as to cause excessively fast cooling of electrons in the afterglow blastwave. 
A possible alternative is a SNR-confined plerion with a globally anisotropic geometry,
 being elongated in the direction of the GRB beam,
 as discussed by KG02 and Guetta \& Granot (2002).
The true efficacy of RT instabilities at the SNR-plerion interface
 should be clarified through future high-resolution numerical simulations
 including all the relevant microphysics.
Arons (2002) has recently discussed in more detail
 the plausibility of a powerful pulsar wind shredding the surrounding SNR
 in the context of a model for cosmic ray acceleration by magnetars.

Although we have concentrated on the case
 of the pulsar spinning down exclusively via a magnetospheric wind,
 there is a possibility that significant spindown
 can also occur through gravitational wave emission,
 such as those driven by r-mode instabilities
 (e.g. Andersson et al. 2000, Stergioulas \& Font 2001, Lindblom, Tohline \& Vallisneri 2001
 and references therein).
However, many aspects of the present theoretical calculations regarding
 gravitational waves are uncertain (Fryer \& Woosley 2001).
Our work can be viewed as exploring the maximum possible effects
 of electromagnetic spindown, which can be modeled with some more confidence.
In reality, both effects could be important to varying degrees,
 and we leave an investigation of such cases to the future.

\acknowledgments

We are thankful to M. Salvati for an initial suggestion leading to the present work,
 and E. Amato, J. Granot, D. Lazzati, N. Omodei and M. Vietri for helpful discussions.
S.I. also expresses his heartfelt thanks to members of the Arcetri Astrophysical Observatory
 for a very pleasant and hospitable environment where part of this work was carried out. 
This work was partly supported by a grant
 from the Italian Ministry of University and Research (Cofin-99-02-02).

\clearpage

 \begin{figure}
 \figurenum{1} 
 \epsscale{0.7} 
 \plotone{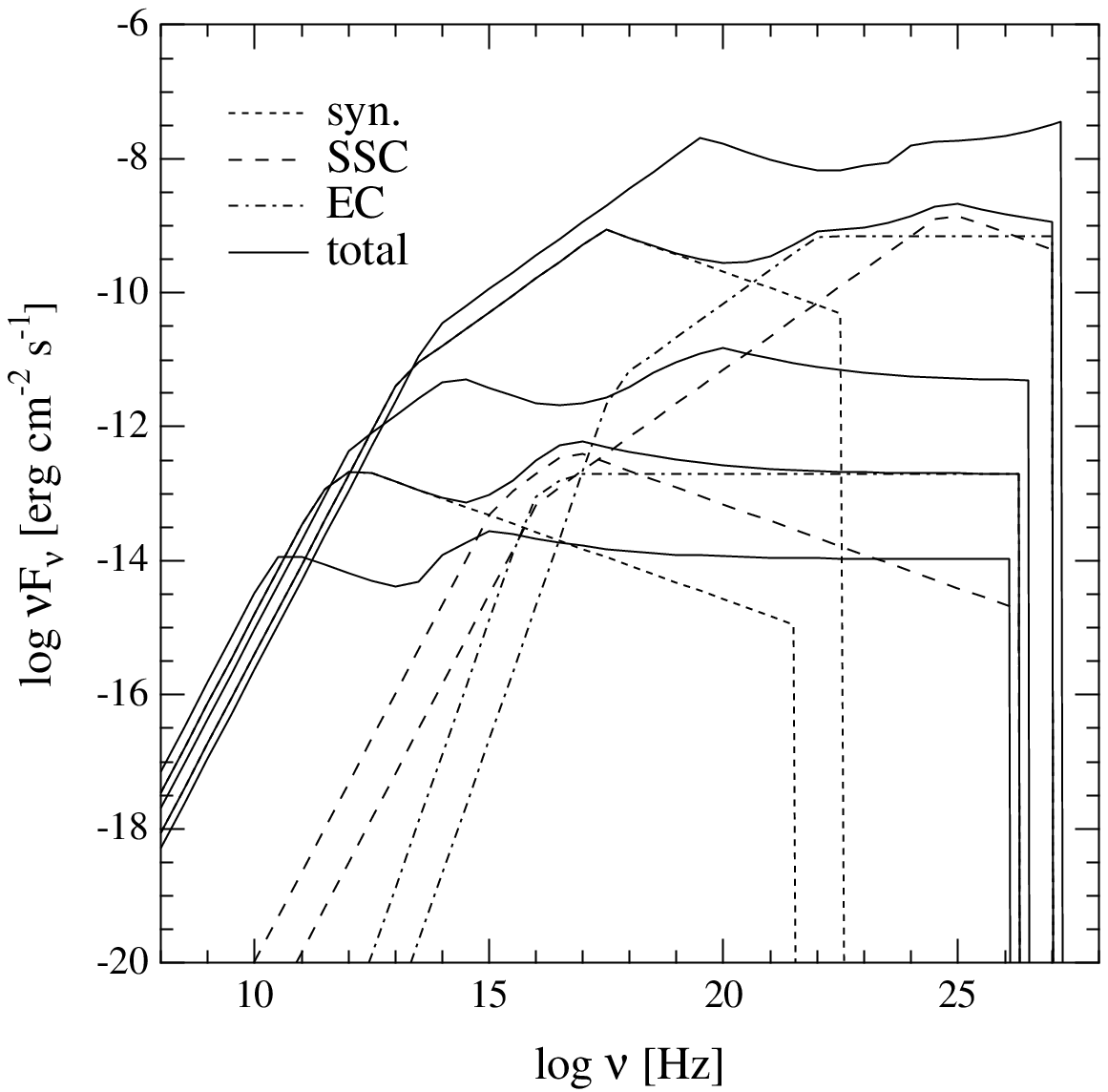} 
 \caption{
The time-dependent afterglow spectra for the fiducial baryon case
 at $t=$1 second, 1 minute, 1 hour, 1 day and 15 days
 after the GRB, from top to bottom.
The curves for $t=$1 minute and $t=$1 day show the decomposition into
 synchrotron (dotted), SSC (dashed) and EC (dot-dashed) components.
} 
 \label{fig:specb}
 \end{figure}

 \begin{figure}
 \figurenum{3} 
 \epsscale{0.7} 
 \plotone{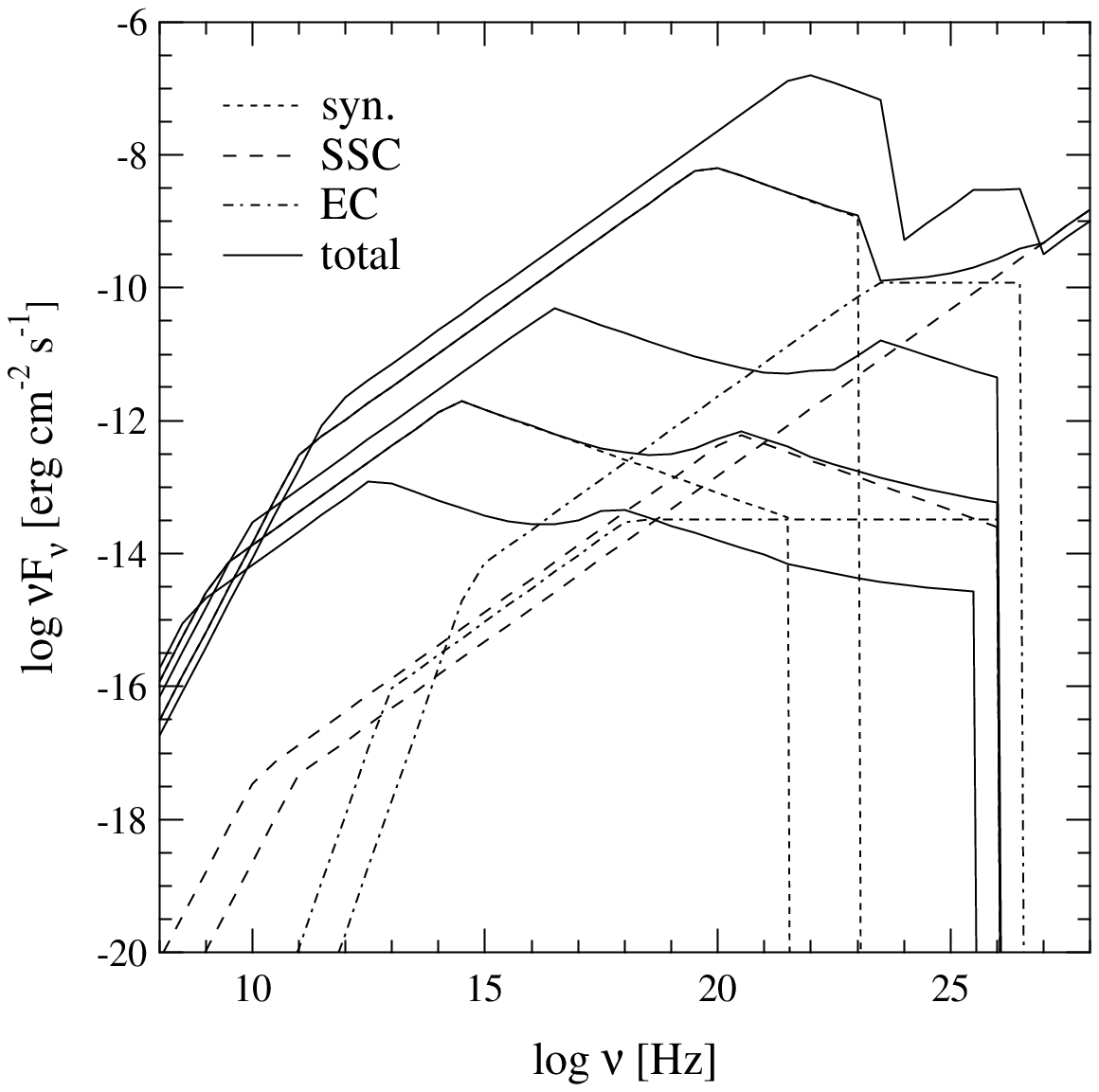} 
 \caption{
As with fig.\ref{fig:specb}, but for the fiducial plerion case.
} 
 \label{fig:specp}
 \end{figure}

\end{document}